\documentstyle[preprint,tighten,aps]{revtex}

\begin{document}
\title{Dynamical study of ${\bf QQ-\bar u \bar d}$ mesons}
 
\author{J. Vijande, A. Valcarce, and K. Tsushima}

\address{Grupo de F\'{\i}sica Nuclear and IUFFyM, Universidad de Salamanca,
E-37008 Salamanca, Spain}
\date{\today}
\maketitle

\begin{abstract}
It has been recently conjectured by Selem and Wilczek \cite{Sel06} the 
existence of a $ss-[\bar u \bar d]$ meson due to strong correlations 
between the two light antiquarks. We make a detailed 
study of this system within a dynamical quark model which 
has proven to be successful in reproducing the most important features 
of low-energy hadron phenomenology. Our results, obtained within a 
parameter-free calculation, show that the antidiquark component of the
$ss \bar u \bar d$ system indeed entails the stronger attraction, 
and drives its energy much lower than 
the $\overline{N}\Xi$ threshold, but still above the 
$\overline{K^0}\,{K^*}^-$ or 
$\overline{{K^*}^0}\,K^-$ thresholds. We have also studied
the $cc \bar u \bar d$ and $bb \bar u \bar d$ systems.
Exotic mesons are only expected to exist in the limit of large 
mass for the two-quark subsystem, $bb\bar u \bar d$, 
since the calculated mass is below the 
$\overline{B^0}\,{B^*}^-$ or
$\overline{{B^*}^0}\,B^-$ thresholds.
\end{abstract}

\vspace*{2cm} 
\noindent Pacs: 12.39.-x, 14.40.-n, 14.65.-q

\newpage

It has been recently re-emphasized \cite{Sel06,Jaf05,Jaf03} 
the potential importance 
of strong diquark (antidiquark) correlations in hadronic 
physics~\cite{Lip71,Kar06}. 
Theoretically the idea of diquark (antidiquark) correlations inside hadrons 
is a consequence of {\it color cancellation}.
The disturbance produced by the color charges of two 
quarks in empty space can be halved by bringing them together 
into a single ${\bf\bar{3}}$ representation of the color $SU(3)$ group.
If this is joined with the more favorable spin-singlet state and Fermi
statistics, the quarks must be in the antisymmetric ${\bf\bar{3}}$ 
representation of flavor $SU(3)$. Besides, one should expect that any
effect disrupting the correlations will induce a repulsive force.
Among such effects we may quote the presence of an additional
diquark or a spectator quark.
Such ideas suggest that the easiest way of constructing low-energy 
exotics could be based on strongly correlated diquarks (antidiquarks)
as building-blocks. 

However, these arguments are rather qualitative, and merely based on 
the group theoretical structure of QCD. 
To study quantitatively whether or not such strong correlations 
between the light quarks (antiquarks) are indeed present,  
one needs QCD-based dynamical studies such as lattice QCD, 
although this eventually must be checked by experiments.
In view of present status of the lattice QCD simulations~\cite{Las06}, 
it is still meaningful to use phenomenological models 
which contain the main features of QCD, once the model parameters 
are calibrated and constrained by as many observables as possible.
Here, we perform such a consistent, parameter-free dynamical 
calculation for the study of the correlations between 
light antiquarks, and investigate the possible existence of 
the exotic meson, $ss-[\bar{u}\bar{d}]$,  
which has been recently conjectured by Selem and Wilczek~\cite{Sel06}.
For completeness, we have also analyzed the
$cc \bar u \bar d$ and $bb \bar u \bar d$ systems.

The present study has been done within the framework of a
constituent quark model which has been successfully applied to study 
the baryon spectra and the baryon-baryon interaction~\cite{Rep05}.
This model has been generalized to include also strange ($s$),
charm ($c$), and beauty ($b$) flavors, and it has also been shown 
to give a reasonable description of the meson spectra \cite{Vij05}.
The description of experimental data gets improved when
four-quark $(qq\bar q \bar q)$ components are also
considered \cite{Vij05b,Vij06}.
The model parameters have been strongly constrained by 
the study of different hadron observables, what represents
an advancing feature compared to studies based on models 
designed {\it adhoc} for a particular problem. 

The model is based on the assumption
that the $u, d$ and $s$ constituent quarks acquire their masses 
due to the spontaneous breaking of the original 
$SU(3)_{L}\otimes SU(3)_{R}$ chiral symmetry at
some momentum scale, which is one of 
the most important nonperturbative phenomena for low energy 
hadron structure.
In this domain of momenta quarks are quasi-particles with constituent
masses interacting through scalar (sigmas, OSE) and pseudoscalar (pions, OPE;
kaons, OKE; and etas, OEE) boson-exchange potentials.
Note that for the case of heavy quarks, $c$ and $b$, 
boson-exchange potentials are not present 
in the model~\cite{Vij06}, since chiral symmetry is badly broken  
already at the level of the current quark masses. 
Beyond the chiral symmetry breaking scale one expects the dynamics
being governed by QCD perturbative effects. They are taken into account
through the one-gluon-exchange (OGE) potential,
a standard color Fermi-Breit interaction.
Finally, any model imitating QCD should incorporate confinement (CON). 
Lattice calculations in the quenched
approximation for heavy quarks show that the confining interaction 
is linearly dependent on the interquark distance. 
The presence of sea quarks, apart
from valence quarks (unquenched approximation), suggests a screening effect on
the potential when increasing the interquark distance. Creation 
of light-quark pairs out of vacuum in between the quarks becomes 
energetically preferable, resulting in a complete screening of 
quark color charges at large distances.
String breaking has been definitively confirmed
through lattice calculations \cite{SESAM}
in coincidence with the quite rapid
crossover from a linear rising to a flat potential well
established in $SU(2)$ Yang-Mills theories \cite{Est99}.
Explicit expressions for the interaction potentials derived from 
the nonrelativistic reduction of the Lagrangian density 
in the static approximation, and a more detailed discussion of the model 
can be found in Ref.~\cite{Vij05}.

For the description of the most general $QQ\bar{u}\bar{d}$ 
($Q=s, c,$ or $b$) system we introduce the Jacobi coordinates, 
\begin{eqnarray}
\text{ \ \ \ \ \ \ \ \ \ \ }\vec{x} &=&\vec{r}_{1}-
\vec{r}_{2},
\text{ \ \ \ \ \ \ \ \ \ \ \ \ \ \ \ }\vec{y
}=\vec{r}_{3}-\vec{r}_{4}, \nonumber \\
\vec{z} &=&\frac{m_{1}\vec{r}_{1}+m_{2}\vec{
r}_{2}}{m_{1}+m_{2}}-\frac{m_{3}\vec{r}_{3}+m_{4}\vec{r
}_{4}}{m_{3}+m_{4}},
\text{ \ \ \ \ \ \ \ \ \ \ \ \ \ }\vec{R}=
\frac{\sum m_{i}\vec{r}_{i}}{\sum m_{i}}, 
\end{eqnarray}
where $1$ and $2$ ($3$ and $4$) stand for quarks (antiquarks). 
The ground state energy of the four-body problem 
can be estimated by a variational method using
a trial wave function that includes all possible color-flavor-spin 
components relevant to a given configuration. For each component,  
$\mid \phi_i>$, such a basis wave function will be a tensor product of
color ($c_i$), flavor ($f_i$), spin ($\chi_i$) and spatial 
($R_i$) parts, 
\begin{equation}
\mid \phi_i>=\mid c_i(1234)> \otimes \mid f_i(1234)> \otimes
\mid \chi _i(1234)> \otimes \mid R_i(1234)>. 
\end{equation}
The most general spatial wave function can be expressed as a
combination of six scalar quantities,
\begin{equation}
\mid R_i(1234)>=R_i(\vec x^{\, 2},\vec y^{\, 2},\vec z^{\, 2},
\vec{x}\cdot\vec{y},\, \vec{x}\cdot\vec{z},\,
\vec{y}\cdot\vec{z}).
\end{equation}
The variational spatial wave function is taken to be 
a linear combination of generalized Gaussians,
\begin{equation}
\mid R_i(1234)>=
\sum_{j=1}^{n} \beta_{i}^{(j)} R_i^{(j)}=
\sum_{j=1}^{n} \beta_{i}^{(j)}
e^{-a^{(j)}_i \vec x^{\,2}-b^{(j)}_i \vec y^{\,2}-c^{(j)}_i \vec 
z^{\,2}-d^{(j)}_i \vec x\cdot \vec y -e^{(j)}_i \vec 
x\cdot \vec z -f^{(j)}_i \vec y\cdot \vec z}, 
\label{wave}
\end{equation}
where $n$ is the number of terms to expand the spatial
wave function of each color-flavor-spin component, and 
$a^{(j)}_i, b^{(j)}_i, ..., f^{(j)}_i$ are the variational parameters.

With respect to the color wave function, $\mid c_i(1234)>$, 
one can couple the two quarks $(1,2)$ and the two antiquarks 
$(3,4)$ to a color singlet state in different ways:

\begin{eqnarray}
&\mid &1_{13},1_{24}> \,\,\, , \,\,\, \mid 8_{13},8_{24}> \,\, ; \label{eq5} \\
&\mid &1_{14},1_{23}> \,\,\, , \,\,\, \mid 8_{14},8_{23}> \,\, ; \label{eq6} \\
&\mid &\overline{3}_{12},3_{34}> \,\,\, , \label{eq7} 
\,\,\, \mid 6_{12},\overline{6}_{34}> \,\, .
\end{eqnarray}
The couplings in Eqs. (\ref{eq5}) and (\ref{eq6}) are convenient for 
asymptotic meson-meson channels (or meson-meson molecules) 
while those in Eq. (\ref{eq7}) are more appropriate
for tetraquark bound states. With our choice of the Jacobi coordinates
the color basis in Eq. (\ref{eq7}) is more suitable 
to deal with the Pauli principle in an easier way.

The spin part can be written as,

\begin{equation}
\mid \chi_i(1234)> =\left[ (12)_{S_{12}}(34)_{S_{34}}\right] _{S},
\end{equation}
where the spin of the two quarks is coupled to $S_{12}$ and that
of the antiquarks to $S_{34}$.

Concerning the flavor part, $\mid f_i(1234)>$, 
since the heavy quarks (those with flavor $s$, $c$ or $b$) have
isospin zero, they do not contribute to the total isospin. 
Therefore one can classify the tetraquark wave function by 
the isospin of the light quarks $I=0,1$.
Taking into account all degrees of freedom, the Pauli principle must be
satisfied for each subsystem of identical quarks (antiquarks). 
It restricts the quantum numbers of the basis 
states, that justifies to use the 
$[(QQ)(\bar u \bar d)]$ coupling.

Using the wave functions described above, we search for a variational solution
for the Hamiltonian. The color, flavor and spin parts are integrated out
and the coefficients $\beta_i^{(j)}$ of the spatial wave function are
obtained by solving the system of linear equations,
\begin{equation}
\sum_i \sum_{j=1}^n \beta_i^{(j)} 
\, [\langle R_{i'}^{(k)}|\,H\,|R_i^{(j)}
\rangle - E\,\langle
R_{i'}^{(k)}|R_i^{(j)}\rangle \delta_{i,i'} ] = 0 
\qquad \qquad {\rm for\,\, all}\,\, k,i', 
\end{equation}
once the eigenvalues $E(a^{(j)}_i, b^{(j)}_i, ..., f^{(j)}_i)$ are obtained
by a minimization procedure with respect to the variational parameters. 
The stable tetraquark states are identified by comparing the 
obtained eigenvalues with the corresponding physical thresholds. 
If they are above the threshold they would be very broad objects, very
hard to detect experimentally.

In a realistic model tetraquarks will not overpopulate the 
meson spectra, in fact they may complement two-quark
components and, indeed, they seem to be necessary
in order to understand the rich meson phenomenology \cite{Vij05b,Vij06}. 
This is due, on the one hand, to the constituent mass
of the quarks, and on the other one, to the finite spectra
generated by screened confining potentials \cite{Vij03}. Only positive 
parity tetraquark states, those that do not need internal orbital
angular momentum between the constituents, may appear in the
low-energy region of the meson spectra and they could mix with $q\bar q $
states with the same quantum numbers. Negative parity four-quark states need a unit of
orbital angular momentum what means an average excitation energy of 800$-$900 MeV \cite{Nir06}.
These ideas have been recently used
to explain the abnormal number of low-energy scalar-isoscalar mesons \cite{Vij05b}
and also the unexpected low masses of positive parity ($0^+$ and $1^+$)
open-charm mesons \cite{Vij06}. They are perfect examples of the way
how the enlargement of the Fock space may help in the understanding
of meson phenomenology. As explained in these works,  
only those states with exotic quantum numbers may appear as
pure four-quark resonances on the meson spectra.
Unfortunately, the present uncertainties on 
the experimental data concerning
exotic channels prevents, for the moment, to extract a definitive
conclusion about its existence \cite{Szc03}.

Let us concentrate on the particular meson state, $ss - [\bar{u}\bar{d}]$,
conjectured in Ref. \cite{Sel06}. It 
has the property of the $[\bar{u}\bar{d}]$ subsystem 
being an {\it antidiquark} state, which means the two antiquarks are in a 
color (${\bf 3}_c$), flavor ($I_{[\bar{u}\bar{d}]}=0$), 
and spin ($S_{[\bar{u}\bar{d}]}=0$) antisymmetric state. 
It requires a completely symmetric radial
wave function for the two antiquarks to satisfy the Fermi statistics.
As antidiquark component $[\bar u\bar d]$ should be in a
relative $S-$wave, one can neglect the crossing terms
in the trial radial wave function, those depending on the
scalar product of different Jacobi coordinates in Eq. (\ref{wave}). 
In order to obtain
a color singlet wave function, the two $s$ quarks must be
in a color antisymmetric, ${\bf \bar 3}_c$, state. Being flavor symmetric,
the corresponding spin wave function may be  either in (i) an antisymmetric, 
$S_{ss}=0$, state that would require the anti-natural radial 
antisymmetric wave function to describe the ground state of the system, 
or in (ii) a symmetric spin state, $S_{ss}=1$, that would combine with 
a natural symmetric radial wave function.
Therefore, the conjectured meson with the presence of the antidiquark 
would be described by a $J^\pi=1^+$ $(L=0,S=1)$ state
with isospin $I=0$. We can summarize the quantum numbers 
of the antidiquark component of the $ss\bar u \bar d$
system in the following way, 
\begin{equation}
|[{\bf 3}_c, S=0, I=0]_{\bar{u}\bar{d}},
|[{\bf \bar 3}_c, S=1, I=0]_{ss}; (S=1,I=0) \rangle.
\label{diqu}
\end{equation}
A full calculation of the $J^\pi=1^+$ $(L=0,S=1)$ state with strangeness $-2$
would require also to consider other vectors in the Hilbert space. In particular,
the same state could also be constructed from a different vector, 
where the two antiquarks would not be an antidiquark state while it still 
has a completely symmetric radial wave function,

\begin{equation}
|[{\bf \bar 6}_c, S=1, I=0]_{\bar{u}\bar{d}},
|[{\bf 6}_c, S=0, I=0]_{ss}, (S=1,I=0) \rangle.
\label{nodiqu}
\end{equation}
This vector, which will be referred to as the {\it nondiquark} component
of the $ss \bar u \bar d$ system, should be considered in the 
calculation of the four-quark state without requiring  
the antidiquark configuration, whereas, it will not be included if only 
the antidiquark configuration is imposed.
For the {\it full} calculation, both the antidiquark and nondiquark 
configurations will be included with the corresponding configuration mixing.

In Table~\ref{t1} we present our results for the antidiquark configuration, 
nondiquark configuration, and full calculation for the 
$ss \bar u \bar d$ system. The same calculation has been
repeated for the $cc \bar u \bar d$ and $bb \bar u \bar d$ systems, 
and the results are presented in Tables~\ref{t2} and \ref{t3}, respectively.

The first important conclusion that can be extracted from the results
of Tables~\ref{t1}, \ref{t2} and \ref{t3} is that the energy of 
the antidiquark configuration is always the lowest. It is interesting
to note how the pseudoscalar force acting between the light quarks 
is responsible for that, 
since the results for the antidiquark and nondiquark configurations 
are almost degenerate if only the confinement and one-gluon exchange 
are retained. The reason for this stems on
the different symmetry for both components in color-spin and 
flavor-spin spaces. While both are symmetric in color-spin space, 
the antidiquark (nondiquark) component is symmetric (antisymmetric) 
in flavor-spin space. 
Therefore, if strong diquark correlations were 
dictated by QCD for light quarks, the dynamical explanation
could not rely on the simple one-gluon exchange dynamics, but it 
would need meson-exchange forces between the constituent quarks.
The similar effect has been also observed in the case of baryon spectra, where
pseudoscalar meson exchanges between the constituent quarks are 
able to revert the relative position in the energy spectra 
of the nucleon Roper resonance, 
with a dominant flavor-spin symmetric wave function, 
with respect to negative parity states, with a 
flavor-spin antisymmetric wave function~\cite{Gar01}. 
It is also interesting to notice that the antidiquark and
nondiquark components are not exactly degenerate when only
the confining interaction is taken into account. This can be easily 
understood by looking at Table~\ref{t4}, where we present 
the contribution of the interaction between $QQ$, $V_{12}$, 
$\bar u\bar d$, $V_{34}$, and $Q \bar n$ ($n=u,d$),
$V_{13}$, for the $QQ\bar u \bar d$ system as a function
of the mass of $Q$, $m_Q$.
The minimization procedure modifies the variational parameter
$a^{(j)}_i$ in Eq. (\ref{wave}) for the $\vec x^{\, 2}$ coordinate 
due to the smaller size of the $QQ$ subsystem
when the mass $m_Q$ increases. As a consequence it gives
a smaller contribution to the energy of the system. In other
words, the dependence on the mass of the quark is introduced
into the calculation through the variational
parameters. 

As predicted by Selem and Wilczek~\cite{Sel06}, the mixing between 
the antidiquark and nondiquark components of the wave function 
diminishes when increasing
the mass of the heavy quarks (see Table \ref{t5}), in such 
a way that for the $b$ quark case
the nondiquark component gives almost no contribution to the
ground state energy of the system. 
This effect, interpreted as the less capacity of the 
spin of a heavy quark to disrupt the correlation of 
the diquark (antidiquark),
comes from the $1/(m_im_j)$ ($m_{i,j}$: constituent quark masses) 
dependence of the one-gluon exchange interaction, which is responsible for 
the mixing between these components. 
Therefore, the mixing decreases with increasing the mass of
the heavy quark.
A similar evidence, the less capacity of the spin of the heavy quarks
to disrupt the system, has also been observed in
the spin-orbit splitting of the
$\Lambda-, \Lambda^+_c-$ and $\Lambda_b-$hypernuclei~\cite{Tsu98},
where $s, c$ and $b$ quarks exclusively carry the
total spin of the $\Lambda, \Lambda^+_c$ and $\Lambda_b$, respectively,
and $u$ and $d$ quarks are coupled to a isospin zero and spin zero
diquark in each baryon.

Regarding the possibility of observing these systems, 
the results obtained are always far
above their corresponding lowest two-meson thresholds, 
as indicated in Tables~\ref{t1}, \ref{t2}, and \ref{t3},
being the only
exception the $bb\bar u\bar d$ system.

Experimentally the possibility to detect a $QQ  \bar u \bar d$ meson 
relies on two different aspects. First of
all the rate of production of $QQ \bar u \bar d$, and second the existence
of decay modes that can provide a unique signature. For the production
at hadronic or $e^+e^-$ colliders one needs to produce two pairs of
charm or bottom quarks. These pairs should be close spatially and the quarks
within each quark-antiquark pair should have small relative momenta in order
to combine in a two-quark, $cc$ or $bb$, state. Finally, these two-quark states
should pick up an antidiquark $[\bar u \bar d]$ to form the desired
$QQ -[\bar u \bar d]$ system. The production rates for the case of charm
quarks have been estimated in Refs. \cite{Gel03} and \cite{Moi96}. The signal
of the strong antidiquark correlation would come from decay channels
preferring to keep the antidiquark structure. So, instead decaying by 
splitting into a two-meson system, it would proceed through a two baryon system
as it would be $\overline{N} \Xi_{cc}$ for the charm case and
$\overline{N} \Xi_{bb}$ for the bottom case. The absence of a dynamical
enhancement of the antidiquark component would open the decay into two
mesons. 
For the $ss \bar u \bar d$ system, if the antidiquark
component is strong enough so as to force it to decay into a two baryon
system via $(ss-[\bar u \bar d]) \to \overline N \Xi$, one can 
expect a natural decay in an $S-$wave, which needs a $J^\pi=1^-$
state for the system. 
As the energy excitation for a unit of orbital angular momentum
costs about 800$-$900 MeV~\cite{Nir06}, this would make the system to be
in the continuum above the $\overline{N} \Xi$ threshold,
therefore being broad and difficult to detect.
The dynamical enhancement of the diquark component is one of the
possible reasons to explain the decay of the $\Lambda(3/2^-)(1520)$
45\% of the time to $N\overline{K}$ channels in order to retain the
diquark structure of the $\Lambda$ inside the $N$. 
Finally, let us remark that we only 
support the possible existence of $QQ\bar u \bar d$ states for $Q=b$ 
and with more uncertainty for $Q=c$, but never for $Q=s$. 
In the bottom case the predicted state should be very narrow and easy to
observe (if produced) since it is far below the 
two-meson and two-baryon physical thresholds.

To summarize, we have made a dynamical, parameter-free, calculation 
for the $QQ\bar u \bar d$ system $(Q=s,c,b)$ within a 
{\it realistic} constituent quark model.
We have found that the antidiquark configuration of these systems always 
gives a lower energy than the nondiquark configuration. 
The mixing between the antidiquark and nondiquark states 
is due to the one-gluon exchange potential, and because of its 
$1/(m_i m_j)$ dependence, it decreases when increasing 
the heavy quark mass.
As a consequence one can expect that the conjectured mesons $QQ-[\bar u
\bar d]$ could be stable for $Q=b$, but we do not find any reason why
these systems should be bound for $Q=s$. Moreover, for light quarks, $Q=s$,
there is no dynamical reason why the antidiquark component should 
be favored compared to the nondiquark one (within an 
uncorrelated quark model). If the existence of any such systems with
two light quarks, $Q=s$, and therefore with a strongly 
correlated light antidiquark would be confirmed via the
postulated baryon-antibaryon final channel, it would mean that a dynamical
mechanism responsible for the correlations has not been considered 
in our simple realizations of models for QCD (based on uncorrelated quarks), 
and the question for the existence of exotic systems, such as 
the pentaquark, should be addressed in a corresponding manner.
On the other hand, the present results may imply that the 
pentaquark with a heavy $\bar{b}$ quark, 
$[u d][u d]\bar{b}$ with a negative parity, may have a chance to be stable,
if the predicted repulsion between diquarks is not strong enough to destroy
the system. This has a merit of further study within the same model. 

\section{Acknowledgements}
This work has been partially funded by Ministerio de Ciencia y 
Tecnolog\'{\i}a under Contract No. FPA2004-05616, by Junta 
de Castilla y Le\'{o}n under Contract No. SA-104/04, and by
Generalitat Valenciana under Contract No.  GV05/276.


\begin{table}
\caption{Calculated energies and physical thresholds, in MeV, for the 
$ss \bar u \bar d$ system. ``Full'' stands for the results 
calculated including both configurations, antidiquark and nondiquark.
Inside the brackets is the percentage of the 
antidiquark component in the full calculation.
\label{tab1}
}
\label{t1}
\begin{center}
\begin{tabular}{|clcccc|}
 & & Antidiquark & Nondiquark & Full & \\
\hline
 & $V_{CON}+V_{OGE}+V_{OPE}+V_{OSE}+V_{OEE}$ & 1705 & 1974 & 1696 (97.59 \%) & \\
 & $V_{CON}+V_{OGE}+V_{OPE}+V_{OSE}$         & 1644 & 1965 &  & \\
 & $V_{CON}+V_{OGE}+V_{OPE}$                 & 1843 & 2105 &  & \\
 & $V_{CON}+V_{OGE}$                         & 2092 & 2083 &  & \\
 & $V_{CON}$                                 & 2520 & 2479 &  & \\
\hline
 &                                           &      &      &  & \\
 & $\overline{N}\, \Xi$ threshold & 2257 & & & \\
 & $\overline{K^0}\,{K^*}^-$ or $\overline{{K^*}^0}\,K^-$ threshold &  & & 1386$-$1390 &\\
\end{tabular}
\end{center}
\end{table}

\begin{table}
\caption{Same as Table~\ref{t1} for the $cc \bar u \bar d$ system.}
\label{t2}
\begin{center}
\begin{tabular}{|clcccc|}
 & & Antidiquark & Nondiquark & Full & \\
\hline
 & $V_{CON}+V_{OGE}+V_{OPE}+V_{OSE}+V_{OEE}$ & 3929 & 4207 & 3927 (99.48 \%) & \\
 & $V_{CON}+V_{OGE}+V_{OPE}+V_{OSE}$         & 3858 & 4210 &  & \\
 & $V_{CON}+V_{OGE}+V_{OPE}$                 & 3906 & 4229 &  & \\
 & $V_{CON}+V_{OGE}$                         & 4169 & 4197 &  & \\
 & $V_{CON}$                                 & 4631 & 4644 &  & \\
\hline
 &                                           &      &      &  & \\
 & $\overline{N}\, \Xi_{cc}$ threshold & 4460 & & & \\
 & $D^+\,{D^*}^0$ or ${D^*}^+\,D^0$ threshold & & & 3875$-$3876 & \\
\end{tabular}
\end{center}
\end{table}

\begin{table}
\caption{Same as Table~\ref{t1} for the $bb \bar u \bar d$ system.} 
\label{t3}
\begin{center}
\begin{tabular}{|clcccc|}
 & & Antidiquark & Nondiquark & Full & \\
\hline
 & $V_{CON}+V_{OGE}+V_{OPE}+V_{OSE}+V_{OEE}$ & 10426 & 10797 & 10426 (99.95 \%) & \\
 & $V_{CON}+V_{OGE}+V_{OPE}+V_{OSE}$         & 10355 & 10801 &  & \\
 & $V_{CON}+V_{OGE}+V_{OPE}$                 & 10403 & 10822 &  & \\
 & $V_{CON}+V_{OGE}$                         & 10673 & 10787 &  & \\
 & $V_{CON}$                                 & 11154 & 11234 &  & \\
\hline
 &                                           &      &      &  & \\
 & $\overline{N}\, \Xi_{bb}$ threshold & & & & \\
 & $\overline{{B^*}^0}\,B^-$ or $\overline{B^0}\,{B^*}^-$ threshold & & & 10604 & \\
\end{tabular}
\end{center}
\end{table}

\begin{table}
\caption{Expectation value, in MeV, of different contributions of the 
confining interaction for the different components of the
$QQ\bar u \bar d$ system and for two different
values of the mass of the quark in the two-quark subsystem.}
\label{t4}
\begin{center}
\begin{tabular}{|cc|cc|ccc|}
 &  &\multicolumn{2}{c|}{$m_Q=555$ MeV} 
 &\multicolumn{2}{c}{$m_Q=5100$ MeV} & \\
 & & Antidiquark & Nondiquark  &  Antidiquark & Nondiquark & \\
\hline
 & $<V_{12}>$ & +458  &  $-$263  & +241 & $-$176 & \\
 & $<V_{34}>$ & +527  &  $-$284  & +513 & $-$264 & \\
 & $<V_{13}>$ & +259  &  +633  & +225 & +536 & \\
\end{tabular}
\end{center}
\end{table}

\begin{table}
\caption{Probability, in \%, of the antidiquark component, $QQ-[\bar u \bar d]$,
as a function of the mass of the quark, in MeV, in the two-quark
subsystem. We also give the mass of the $QQ\bar u \bar d$ system in MeV.}
\label{t5}
\begin{center}
\begin{tabular}{|ccccc|}
 & $m_Q$ & $M_{QQ\bar u \bar d}$ & P($QQ-[\bar u \bar d]$) &  \\
\hline
 & 313  &  1431   & 96.09 &\\
 & 555  &  1696   & 97.59 &\\
 & 755  &  1980   & 98.18 &\\
 &1255  &  2815   & 99.11 &\\
 &1555  &  3352   & 99.37 &\\
\end{tabular}
\end{center}
\end{table}

\end{document}